\documentclass[12pt]{article}
\usepackage{cite}

\newcommand{\beq}{\begin{equation}}
\newcommand{\eeq}{\end{equation}}
\newcommand{\bea}{\begin{eqnarray}}
\newcommand{\eea}{\end{eqnarray}}
\newcommand{\bean}{\begin{eqnarray*}}
\newcommand{\eean}{\end{eqnarray*}}
\newcommand{\pder}[2]{\frac{\partial {#1}}{\partial {#2}}}
\newcommand{\deriv}[2]{\frac{{\rm d} {#1}}{{\rm d} {#2}}}
\newcommand{\ave}[1]{\langle #1 \rangle}
\newcommand{\steady}[1]{{\overline{#1}}}
\newcommand{\sech}{{\rm \, sech}}
\newcommand{\cosech}{{\rm \, cosech}}

\renewcommand{\vec}[1]{\mbox{\boldmath $#1$}}

\newcommand{\tsfrac}[2]{{\textstyle\frac{#1}{#2}}}
\newcommand{\dsfrac}[2]{{\displaystyle\frac{#1}{#2}}}
\newcommand{\e}{{\mathrm e}}
\newcommand{\dam}{{\mathrm{Da}}}

\newcommand{\Ord}{{\mathcal O}}

\usepackage[dvips]{epsfig}
\title{A bistable reaction--diffusion system in a stretching flow} 

\author{Stephen M. Cox${}^1$ and Georg A. Gottwald${}^2$\\
\small
${}^1$School of Mathematical Sciences, University of Adelaide, Adelaide 5005,
Australia\\
\small
${}^2$School of Mathematics and Statistics, University of Sydney, NSW 2006,
Australia}

\begin{document}

\maketitle

\begin{abstract}

We examine the evolution of a bistable reaction in a one-dimensional
stretching flow, as a model for chaotic advection. We derive two reduced
systems of ordinary differential equations (ODE's) for the dynamics of the
governing advection--reaction--diffusion partial differential equation
(PDE), for pulse-like and for plateau-like solutions, based on a
non-perturbative approach. This reduction allows us to study the dynamics
in two cases: first, close to a saddle--node bifurcation at which a pair
of nontrivial steady states are born as the dimensionless reaction rate
(Damk\"ohler number) is increased, and, second, for large Damk\"ohler
number, far away from the bifurcation. The main aim is to investigate the
initial-value problem and to determine when an initial condition subject
to chaotic stirring will decay to zero and when it will give rise to a
nonzero final state. Comparisons with full PDE simulations show that the
reduced pulse model accurately predicts the threshold amplitude for a pulse
initial condition to give rise to a nontrivial final steady state, and
that the reduced plateau model gives an accurate picture of the dynamics of the
system at large Damk\"ohler number.

\end{abstract}



\clearpage

\section{Introduction} In natural and industrial environments active
processes such as chemical or biological reactions are often embedded in a
fluid flow. Examples include mixing of reactants within continuously fed
or batch reactors \cite{Epstein95,Allen96}, the development of plankton
blooms and occurrence of plankton patchiness
\cite{Abraham98,Martin00,McLeod02,HG04}, increased
depletion of ozone caused by chlorine filaments \cite{Edouard96} and
flame filamental structures in combustion systems~\cite{Ronney94}. The
fluid flow is very often time-dependent and stirring. This gives rise
to interesting effects, and the presence of compression and stirring
in chaotic flows typically leads to filamentation of the chemical or
biological components. This evidently changes the reaction
dynamics. The filamentation caused by the chaotic advection leads to
an increased surface area of the reacting components and an increase
of the reaction output. However, if the stirring rate is too strong,
the filaments become thinner and thinner and the reaction may
stop. The corresponding flow-mediated saddle--node bifurcation has
been reported in
\cite{Neufeld01,Neufeld02,Neufeld02b,HG03,Kiss03b,Kiss03,Menon05,Ranz79}.

A natural question in such stirred reactions is: Given a chaotic flow
environment and an initial distribution of chemical or biological
reactants, is it possible to predict whether the reaction will take
place and develop or whether it will die out? This question is of
paramount importance in an environmental context where the reactants
may be ozone in the atmosphere or pollutants in the ocean, or in an
industrial context where one is interested in maximizing the reaction
output. This problem is well-studied in the non-stirred case~\cite{Gray94}.
However, so far the complex nature of chaotic flows has
prohibited an analytical treatment of this problem in the stirred case.

In principle, these phenomena can be studied directly in a two- or
three-dimensional reaction--advection--diffusion system, albeit with huge
computational effort. An analytical treatment of the full system is
prohibited by the complicated nature of the underlying equations, which
involve multiple-scale processes. Simplified models are needed to capture
essential features of the influence of the stirring on the reaction kinetics.
In filamental or lamellar models~\cite{r79}, the two-dimensional problem
of reacting tracers is replaced by a one-dimensional problem of the form
\beq
\label{gen_eqn}
\frac{\partial}{\partial t} u_i - \lambda x \frac{\partial}{\partial
x} u_i = D_i \frac{\partial^2}{\partial x^2} u_i + {\cal{F}}_i(u_i;k_i) \; ,
\qquad i=1,\ldots,n,
\eeq
for $n$ reacting tracers ($u_i$, $i=1,\ldots,n$) with diffusion
coefficients $D_i$, reaction rates $k_i$ and stirring rate $\lambda$. Such
models have been applied by several groups, to autocatalytic, bistable and
excitable media in several physical, chemical and biological contexts
\cite{Martin00,McLeod02,Neufeld01,Neufeld02,Neufeld02b,HG03,Kiss03b,Kiss03}.
Phenomenological filamental models such as (\ref{gen_eqn}) can be
justified by the following consideration: The chaotic advection causes
filaments to be stretched in one direction and compressed in another. In
the stretched direction, the concentration is homogenized and gradients
along the filaments can be neglected. This motivates a one-dimensional
reduction for the concentration in the direction transverse to the
filament, subject to the effect of stirring and compression. The parameter
$\lambda$ can be thought of as the Lagrangian mean strain in the
contracting direction, and may be argued to be given by the absolute value
of the negative Lyapunov exponent or the (slightly larger) topological
entropy. For a different approach to this problem see \cite{Boozer99,JLT03}.
The validity of such simplified models has been numerically investigated in
\cite{Cox04}. 

In this paper we study a bistable model with $n=1$ and
${\cal{F}}_1(u_1;k_1)=k_1u_1(u_1-1)(\alpha-u_1)$ in (\ref{gen_eqn}). The
solutions of this model are found to be generically either of a plateau-like
or a pulse-like character. Plateau solutions are found stably 
when the stirring rate is low compared to the reaction rate. Pulse-like
solutions are found close to the saddle--node bifurcation, where the
stirring rate is strong enough to suppress the reaction; moreover the
unstable solution for weak stirring rates is also of a pulse-like
character. The unstable solution will be of interest when we consider the
initial-value problem and ask when an initial perturbation will develop
into a plateau-like solution. We extend a non-perturbative approach
developed in \cite{GottwaldKramer04} to derive from (\ref{gen_eqn}) a
corresponding low-dimensional system of ordinary differential equations
which describes the time evolution of plateau-like and
pulse-like solutions. We determine the equilibrium solutions, 
accurately describe the saddle--node behaviour and derive an analytical
expression for the asymptotic width of a stationary front. By studying the
phase-portrait of the reduced system we are able to determine the fate of
various initial concentration profiles in (\ref{gen_eqn}); these
predictions are then verified numerically by comparison with simulations
of the full partial differential equation (\ref{gen_eqn}). 

In Section \ref{Sec_model} we introduce the bistable model and discuss its
solutions and their bifurcations. In Section \ref{Sec_varia} we describe
the non-perturbative variational approach. This method will be used in
Section \ref{Sec_pulses} to describe pulse-like solutions and in Section
\ref{Sec_fronts} to describe plateau-like solutions. In Section
\ref{Sec_initial} we use the reduced system derived in Sections
\ref{Sec_pulses} and \ref{Sec_fronts} to characterise the initial-value
problem. The paper concludes with a discussion and an outlook for further
research. 


\section{The model}
\label{Sec_model}

In this paper we study a bistable model 
\begin{equation}
\label{eq:ard}
\frac{\partial}{\partial t} u - x \frac{\partial}{\partial
x} u = D \frac{\partial^2}{\partial x^2} u + \dam\, u(u-1)(\alpha-u),\qquad
\mbox{$u(x,t)\to0$ as $x\to\pm\infty$} \; ,
\end{equation}
where $0<\alpha<1$. The Damk\"{o}hler number $\dam=k/\lambda$ measures the
ratio of the time scales of fluid motion and reaction: small Damk\"{o}hler
numbers correspond to fast stirring/slow reaction, while for large
Damk\"{o}hler numbers the system behaves asymptotically like an unstirred
system. 

The ODE corresponding to (\ref{eq:ard}) in the absence of spatial
effects has two stable fixed points, $u=0$ and $u=1$, which are
separated by an unstable fixed point at $u=\alpha$. For the unstirred
case of the PDE (\ref{eq:ard}), the system is well known and well
described in textbooks such as \cite{Murray,Keener,Mikhailov,Scott}:
an initial perturbation which is larger than $\alpha$ over a finite
range will spread over the whole domain if $0<\alpha<0.5$; by
contrast, if $0.5<\alpha<1$, an initial perturbation will decay to the
stable state $u=0$. The stirred case is much less well understood, and
the initial-value problem has to our knowledge never been analytically
treated. The stirred case was investigated numerically in
\cite{McLeod02,Neufeld02}, and semi-analytically in \cite{Menon05}:
stationary fronts between the $u=0$ and $u=1$ states exist as a
balance between the $x$-dependent stirring and the diffusion-mediated
counterpropagating fronts, for large enough values of the
Damk\"{o}hler number. An initial sufficiently large perturbation
seeded at $x=0$ spreads as a pair of fronts, driven by its reaction
kinetics and diffusion, until the fronts reach the locations $x=\pm
\nu$ where their velocities equal the ambient spatially dependent
velocity of the chaotic stirring.

It has been observed in \cite{Neufeld02,Menon05} that there is a critical
Damk\"{o}hler number such that no stationary pulses exist for
$\dam<\dam_c$, i.e., when the time scale of the chaotic advection,
$\tau_f=1/\lambda$, becomes too short in comparison with the time scale of
the reaction, $t_r=1/k$. For large Damk\"{o}hler numbers, an asymptotic
expression for the scaling of the total concentration was developed in
\cite{Neufeld02}. 
However, the techniques used in \cite{Neufeld02} cannot describe the
behaviour close to the bifurcation point $\dam=\dam_c$.  Moreover,
these techniques are stationary and cannot answer questions about the
evolution of initial perturbations of reactants.

\begin{figure}
\begin{center}
\epsfig{file=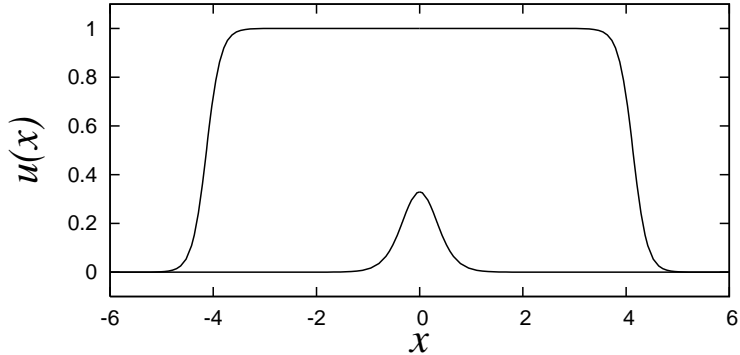,width=0.6\linewidth}
\end{center}
\caption{The two nontrivial steady states of (\ref{eq:ard}) for
$\alpha=0.2$ and $\dam=100$. The `pulse' solution is unstable; the `plateau'
solution is stable.}
\label{fig:intro}
\end{figure}

We now describe in more detail the asymptotic form of the two nontrivial
steady states for large $\dam$.

\subsection{Asymptotic steady states for large $\dam$}
\label{Sec_asympt}

We aim to find steady states $u(x)$ that satisfy
\beq
xu_x+Du_{xx}+\dam\,u(\alpha-u)(u-1)=0,\qquad
\mbox{$u(x)\to0$ as $x\to\pm\infty$},
\label{eq:ss}
\eeq
when $\dam\gg1$.
Numerical simulations show that for $\dam \gg1$ the stable solution is
plateau-like and the unstable solution is bell-shaped
(see Figure~\ref{fig:intro}).

For the unstable bell-shaped solution $u=V(x)$, we introduce
\[
X=\sqrt{\dam}\,x
\]
so that (\ref{eq:ss}) becomes
\[
DV_{XX}+V(\alpha-V)(V-1)=0,
\]
with an error of order $\dam^{-1}$.  We may now readily determine the
appropriate leading-order solution, which gives rise to
\beq
V(x)=3\alpha\left\{1+\alpha+\sqrt{\tsfrac12(1-2\alpha)(2-\alpha)}
\cosh\left(\sqrt{\frac{\alpha\dam}{D}}x\right)\right\}^{-1}
+\Ord(\dam^{-1}).
\label{eq:exact1}
\eeq
We note that this gives
\beq
V(0)
\sim3\alpha\left(1+\alpha+\sqrt{\tsfrac12(1-2\alpha)(2-\alpha)}\right)^{-1},
\label{eq:V(0)}
\eeq
which we shall later compare with a corresponding result obtained from
a reduced test function ansatz using a bell-shaped test function,
derived in Section~\ref{Sec_pulses}.

In the large-$\dam$ limit, the plateau solution $u=U(x)$ is, to leading
order in $\dam$, given by $U=1$ in a region $-\nu <x< \nu$ and $U=0$
outside this region, where $\nu=\sqrt{\dam}\,\omega$, for some
$\omega=\Ord(1)$. 
The locations of the interfaces are determined by a balance between the
tendency of the pulse to spread (with constant velocity) and the compressive
effects of the imposed velocity field (for which the velocity is proportional
to $x$)~\cite{Neufeld02,Menon05}; we return to this point in more detail
in Section~\ref{sec:5.1.1}.  Around $x=\pm \nu$ there are transition regions of
width $\Ord(1/\sqrt{\dam})$. To analyse the region around $x=\nu$, we
introduce
\[
\xi=\sqrt{\dam}(x-\sqrt{\dam}\,\omega).
\]
Then from (\ref{eq:ss}) we obtain
\[
DU_{\xi\xi}+\omega U_\xi+U(\alpha-U)(U-1)=0,
\]
with an error of order $1/\sqrt{\dam}$.
Correspondingly,
\beq
U(\xi)=\tsfrac12\left(1-\tanh\left[\xi/\sqrt{8D}\right]\right),
\label{eq:exact2}
\eeq
with $\omega=\sqrt{D/2}(1-2\alpha)$.  There is a similar transition
region near $x=-\nu$. We shall in Section~\ref{Sec_fronts} use $\tanh$
profiles such as these as an ansatz for our second reduced model;
since the ansatz captures exactly the large-$\dam$ form of the fronts,
we shall find correspondingly that the reduced model provides an
excellent approximation to the dynamics of the full PDE
(\ref{eq:ard}).

Figure \ref{fig:intro} shows the solutions of (\ref{eq:ard}) obtained by a
shooting algorithm for high Damk\"ohler number ($\dam=100$). It clearly
shows that the stable solution is plateau-like and the unstable solution is
pulse-like. Note that the stable solution for smaller $\dam$, close to the
bifurcation (not shown here), is also pulse-like. 

\subsection{Asymptotic solutions for small $\dam$}
\label{Sec_asympt2}

\begin{figure}
\begin{center}
\epsfig{file=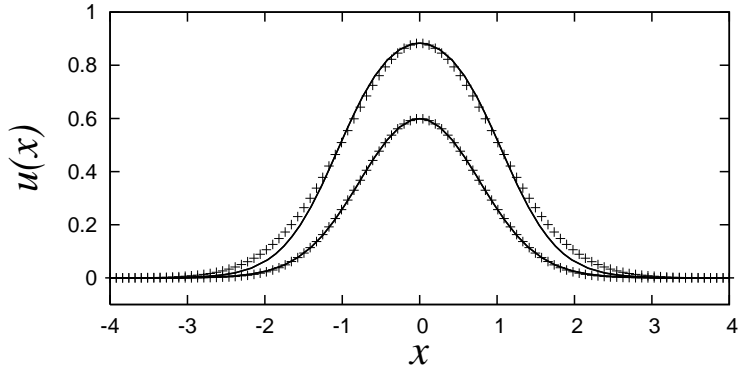,width=0.6\linewidth}
\end{center}
\caption{The two nontrivial steady states of (\ref{eq:ard}) for
$\alpha=0.2$ and $\dam=10$; the upper and lower solutions are, respectively,
stable and unstable. Shown in each case with crosses are fits to a Gaussian.}
\label{fig:intro2}
\end{figure}

We now seek an appropriate ansatz for the solution behaviour close to the
saddle--node bifurcation. Some insight into the form of the solution may
be gained by examining the small Damk\"ohler number limit of the full
time-dependent problem (\ref{eq:ard}). Beyond the saddle--node bifurcation
point, i.e., for $\dam < \dam_c$, no steady-state solutions exist, and
any initial condition decays to zero. In fact, if we expand the
solution as a series in the Damk\"ohler number according to $u(x,t)=\dam\,
u_0(x,t)+{\cal{O}}(\dam^2)$, the equation is satisfied at leading order by
\begin{equation}
u_0(x,t)=f_0(t)\exp(-(w_0(t)x)^2)\; ,
\label{eq:smallDa0}
\end{equation}
where $w_0\to 1/{\sqrt{2D}}$ and $f_0 \to 0$ as $t\to\infty$; the latter
limit reflects the nonexistence of nontrivial steady states for $\dam<\dam_c$.
Although $\dam_c=\Ord(10)$, which is clearly not small,
it seems that the leading-order term 
(\ref{eq:smallDa0}) provides an accurate representation of the solution
to (\ref{eq:ard}) close to the saddle--node point $\dam=\dam_c$
-- see Figure~\ref{fig:intro2}.
So in the first instance we shall approximate both stable and unstable
solutions close to the saddle--node point by a Gaussian; we shall later
consider, in somewhat less detail, other possible profiles of `bell-shaped'
form.

Using a non-perturbative method developed in \cite{GottwaldKramer04}
we now derive a set of ordinary differential equations which describe the
dynamics of these plateau- and pulse-like solutions. In the next section
we briefly explain this variational approach.


\section{Nonperturbative, variational method}
\label{Sec_varia}

A method was developed in \cite{GottwaldKramer04} to study critical wave
propagation of single pulses and pulse trains in excitable media in one
and two dimensions. It was based on the observation that close to the
bifurcation point the pulse shape is approximately a bell-shaped function.
Numerical simulations and the asymptotic analysis in Section
\ref{Sec_asympt} and \ref{Sec_asympt2} show that this is also the case
for the bistable model (\ref{eq:ard}) close to the saddle--node
bifurcation at which the stable and unstable solutions are born, and
on the unstable branch at large Damk\"ohler numbers.
A test-function approximation that optimises the
two free parameters of a bell-shaped function, i.e., its amplitude and
its width, allows us to find the actual bifurcation point, $\dam_c$,
and determine the pulse shape for close-to-critical pulses at
Damk\"ohler numbers near $\dam_c$. We note that the framework of
asymptotic techniques, such as inner and outer expansions where the
solution is separated into a steep narrow front and a flat plateau,
are bound to fail close to the bifurcation point, because the pulse is
clearly bell-shaped, and such a separation is no longer possible.  We
shall make explicit use of the shape of the pulse close to the
critical point and parameterise the pulse appropriately.

We choose $u(x,t)$ of the general form
\beq
\label{ansatz}
u(x)=f(t) U(\eta) \qquad {\rm{with}} \qquad \eta= w(t) x \; ,
\eeq
where $U(\eta)$ is a symmetric, bell-shaped function 
of unit width and height, and $f(t)$ is the amplitude of the
pulse. Close to the saddle--node bifurcation and for the unstable solution
at large Damk\"ohler numbers, the solution is indeed a bell-shaped pulse,
which we approximate by a Gaussian (see Figure~\ref{fig:intro2}).
However, our results do not depend on
the specific choice of the test function, and the numerical values differ
only marginally when $\sech$-functions are used instead, for instance
(see Figure~\ref{fig:pulse-sech2}). We
restrict the solutions to a subspace of a bell-shaped function $fU(\eta)$,
which is parameterised by the amplitude, $f(t)$, and the inverse pulse
width, $w(t)$. The evolution of these parameters is then determined by
minimizing the error made by the restriction to the subspace defined by
(\ref{ansatz}); this is achieved by projecting (\ref{eq:ard}) onto the
tangent space of the restricted subspace, which is spanned by $\partial u
/ \partial {f}=U$ and $\partial u / \partial w=f \eta U_{\eta}/w$.
We set to zero the
integral of the product of (\ref{eq:ard}) with the basis functions of the
tangent space (over the entire $\eta$-domain). This leads to ordinary
differential equations for $f(t)$ and $w(t)$; it also yields an
approximation to the critical Damk\"ohler number $\dam_c$. 

One may in fact choose any plausible test function and optimize the
restriction to the particular space of solutions by varying the
parameters. This variational idea is not restricted to pulse-like
solutions. Far away from the saddle--node bifurcation, at large
Damk\"ohler numbers, where the stable asymptotic solution comprises a pair of
fronts (see Section \ref{Sec_asympt}) we may use as a test function a
combination of $\tanh$-functions
\begin{equation}
u(x,t)=\tsfrac12 f(t)\left[\tanh(\eta+a(t))-\tanh(\eta-a(t))\right]\qquad
\mbox{with}\qquad \eta=w(t)x,
\label{eq:utanh}
\end{equation}
where $a(t)=w(t)\nu(t)$ and the fronts are located around $x=\pm\nu$.  In a
manner analogous to that described above for a pulse ansatz, the
variational approach allows us in this case to determine the time
evolution of the solution amplitude $f(t)$, the inverse interface width
$w(t)$ and the front locations $\pm\nu(t)$
by projecting (\ref{eq:ard}) onto the tangent space of the restricted
solution subspace, which is now spanned by $\partial u / \partial
{f}$, $\partial u / \partial w$ and $\partial u / \partial \nu$.


\section{Stable and unstable pulse solutions}
\label{Sec_pulses}

Near the turning point at $\dam=\dam_c(\alpha)$, both nontrivial
steady-state solutions take the form of bell-shaped pulses near
$x=0$. We note that in the literature a {\it pulse} commonly refers
to a stable homoclinic solution; however, here we use the term merely
to mean a bell-shaped function, regardless of whether it is stable
or unstable. This terminology allows us to distinguish between the two
test function ans{\"a}tze (\ref{ansatz}) and (\ref{eq:utanh}).

To determine a reduced model for such pulse solutions, we
therefore introduce the ansatz (\ref{ansatz})
and to determine the evolution of $f(t)$ and $w(t)$ we enforce
the vanishing of the two inner products
\bean
\langle (u_t-xu_x-Du_{xx}-\dam \, u(\alpha-u)(u-1)) u_f\rangle&=&0\\
\langle (u_t-xu_x-Du_{xx}-\dam \, u(\alpha-u)(u-1)) u_w\rangle&=&0,
\eean
where
\[
u_f\equiv\pder{u}{f}=U,\qquad
u_w\equiv\pder{u}{w}=\frac{f}{w}\eta U',\qquad
\langle \cdots \rangle \equiv\int_{-\infty}^\infty \cdots\,{\rm d}\eta.
\]

For a Gaussian pulse, with $U(\eta)=\e^{-\eta^2}$, we note the
requisite integrals
\[
\ave{U^m}=\left(\frac{\pi}{m}\right)^{1/2},\qquad
\ave{{U'}^2}=\left(\frac{\pi}{2}\right)^{1/2},\qquad
\ave{\eta^2{U'}^2}=\frac{3}{4}\left(\frac{\pi}{2}\right)^{1/2}.
\]
Thus after some algebra we find that $f(t)$ and $w(t)$ evolve according to
\bea
\dot{f}&=&f
\left\{-\dam\alpha
-2Dw^2+\tsfrac73\dam(1+\alpha)\sqrt{\tsfrac16}f
-\tsfrac54\dam\sqrt{\tsfrac12}f^2
\right\}\label{eq:fd}\\
\dot{w}&=&w
\left\{
1-2Dw^2+\tsfrac13\dam(1+\alpha)\sqrt{\tsfrac23}f
-\tsfrac12\dam\sqrt{\tsfrac12}f^2
\right\}.\label{eq:wd}
\eea
In order to examine the dynamics of this system in the phase plane it is
useful for brevity of notation to write it in the form
\bea
\dot{f}&=&f\left\{-\mu_1-\mu_2w^2+\mu_3f-\mu_4f^2\right\}\label{eq:fdp}\\
\dot{w}&=&w\left\{\lambda_1-\lambda_2w^2+\lambda_3f-\lambda_4f^2\right\},
\label{eq:wdp}
\eea
and note that the $\mu_i$ and $\lambda_i$ are all positive; also
$\mu_2=\lambda_2$ (this equality does not necessarily hold for other
profiles $U(\eta)$).

We now examine the steady states $(w,f)=(\steady{w},\steady{f})$ of
(\ref{eq:fdp}), (\ref{eq:wdp}) and their stability. Note that the
steady-state problem has been considered previously \cite{Menon05},
although only the solutions identified below as `$P_4$' were considered
there. Although not all steady states of (\ref{eq:fdp}), (\ref{eq:wdp})
correspond to distinct physical solutions, they help us to map out the
structure of the phase plane, and hence determine the dynamics of
(\ref{eq:fdp}), (\ref{eq:wdp}). 

\subsubsection*{$P_1$: $(\steady{w},\steady{f})=(0,0)$}

This steady state exists for all parameter values, and a linearisation
about $P_1$ yields the eigenvalues $\lambda_1$ and $-\mu_1$. Therefore
$P_1$ is a saddle point, unstable to perturbations in $w$ and stable to
perturbations in $f$.

\subsubsection*{$P_2$:
$(\steady{w},\steady{f})=((\lambda_1/\lambda_2)^{1/2},0)$}

This steady state exists for all parameter values. A linearisation about this
point yields the eigenvalues $-\mu_1-\mu_2\lambda_1/\lambda_2<0$ and
$-2\lambda_1<0$. Thus $P_2$ is a stable node.

\subsubsection*{$P_3$: $\steady{w}=0$, $\steady{f}\neq0$}

Here
\beq
\mu_4\steady{f}^2-\mu_3\steady{f}+\mu_1=0.
\label{eq:P3}
\eeq
This quadratic has two real, positive solutions 
\bean
\steady{f}_-&=&\frac{\mu_3-(\mu_3^2-4\mu_4\mu_1)^{1/2}}{2\mu_4}\\
\steady{f}_+&=&\frac{\mu_3+(\mu_3^2-4\mu_4\mu_1)^{1/2}}{2\mu_4},
\eean
for all parameter values; indeed $\steady{f}_-$ and $\steady{f}_+$ are
independent of $\dam$. Note that, since $P_3$ corresponds to
infinitely wide pulses ($\steady{w}=0$), the roots of the quadratic
would be $\alpha$ and $1$ if there were no approximation due to the
Gaussian pulse ansatz.  A linearisation about either steady state
yields the eigenvalues
\[
e_1\equiv (\mu_3-2\mu_4\steady{f})\steady{f},\qquad
e_2\equiv (\lambda_1+\lambda_3\steady{f}-\lambda_4\steady{f}^2).
\]
Now $\mu_3-2\mu_4\steady{f}_\pm=\mp(\mu_3^2-4\mu_4\mu_1)^{1/2}$, so
the eigenvalue $e_1$ indicates stability for $f_+$ and instability for
$f_-$ ($e_1$ corresponds to perturbations in $f$, with $w\equiv0$).
The eigenvalue $e_2$ corresponds to disturbances in $w$ and satisfies
\[
\mu_4 e_2=(\lambda_3\mu_4-\lambda_4\mu_3)\steady{f}+
          (\lambda_4\mu_1+\lambda_1\mu_4);
\]
it seems to indicate instability for $f_+$ and either stability or 
instability for $f_-$, depending on the parameter values.

\subsubsection*{$P_4$: $\steady{w}\neq0$, $\steady{f}\neq0$}

Here $\steady{f}$ satisfies
\beq
(\mu_4-\lambda_4)\steady{f}^2-(\mu_3-\lambda_3)\steady{f}+\mu_1+\lambda_1=0.
\label{eq:P4}
\eeq
We note the coefficients (cf.~\cite{Menon05})
\beq
\mu_4-\lambda_4=\frac{3\dam}{4\sqrt{2}},\qquad
\mu_3-\lambda_3=\frac{5\dam(1+\alpha)}{3\sqrt{6}},\qquad
\mu_1+\lambda_1=1+\dam \, \alpha.
\label{pulse_f0}
\eeq
The quadratic (\ref{eq:P4}) has two real, positive roots provided
\[
\alpha<\alpha_m\equiv\frac{1-2q-\sqrt{1-4q}}{2q}\approx0.47448,\qquad
\mbox{where }q=\frac{25}{81\sqrt{2}},
\]
and
\bean
\dam>\dam_c^{\rm pulse}\equiv\left[q(1+\alpha)^2-\alpha\right]^{-1}.
\eean
Note that $\alpha_m$ is slightly smaller than $0.5$, the threshold
for existence of non-zero solutions in the unstirred case.

In this case, we denote the two solutions of (\ref{eq:P4}) by $f^*_+>f^*_-$.
Correspondingly $\steady{w}=w^*_\pm$, where
\beq
\mu_2{w^*_\pm}^2=-\mu_4{f^*_\pm}^2+\mu_3f^*_\pm-\mu_1.
\label{eq:mu2w2}
\eeq
Since the right-hand side of (\ref{eq:mu2w2}) must be non-negative,
we deduce, by comparison with (\ref{eq:P3}), that each of $f^*_-$
and $f^*_+$ lies between $f_-$ and $f_+$.
At $\dam=\dam_c^{\rm pulse}$, the two solutions collide in a saddle--node
bifurcation; no such solutions exist for $\dam<\dam_c^{\rm pulse}$.

A linearisation about either of the $P_4$ steady states yields, for
$f=\steady{f}+\delta f(t)$ and $w=\steady{w}+\delta w(t)$, that
\bean
\delta \dot{f}&=&(\mu_3\steady{f}-2\mu_4\steady{f}^2)\,\delta f
-2\mu_2\steady{w}\steady{f}\,\delta w\\
\delta \dot{w}&=&(\lambda_3-2\lambda_4\steady{f})\steady{w}\,\delta f
-2\lambda_2\steady{w}^2\,\delta w.
\eean
Hence the growth rate $\sigma$ satisfies
\beq
\sigma^2+b\sigma+c=0,
\label{sigma}
\eeq
where
\[
b=2\lambda_2\steady{w}^2-\mu_3\steady{f}+2\mu_4\steady{f}^2
=-2\mu_1+\mu_3\steady{f}
\]
and
\[
c=2\mu_2\steady{w}^2\steady{f}(2(\mu_4-\lambda_4)\steady{f}-(\mu_3-\lambda_3)).
\]
Note that for $f^*_+$, $c>0$; for $f^*_-$, $c<0$, indicating
instability.  The bifurcation point $c=0$ corresponds to (the limits
$\steady{w}=0$ or $\steady{f}=0$ or) $\dam=\dam_c^{\rm pulse}$. Note
that $\sigma=\Ord(\dam)$ as $\dam\to\infty$.

\begin{figure}
\begin{center}
\epsfig{file=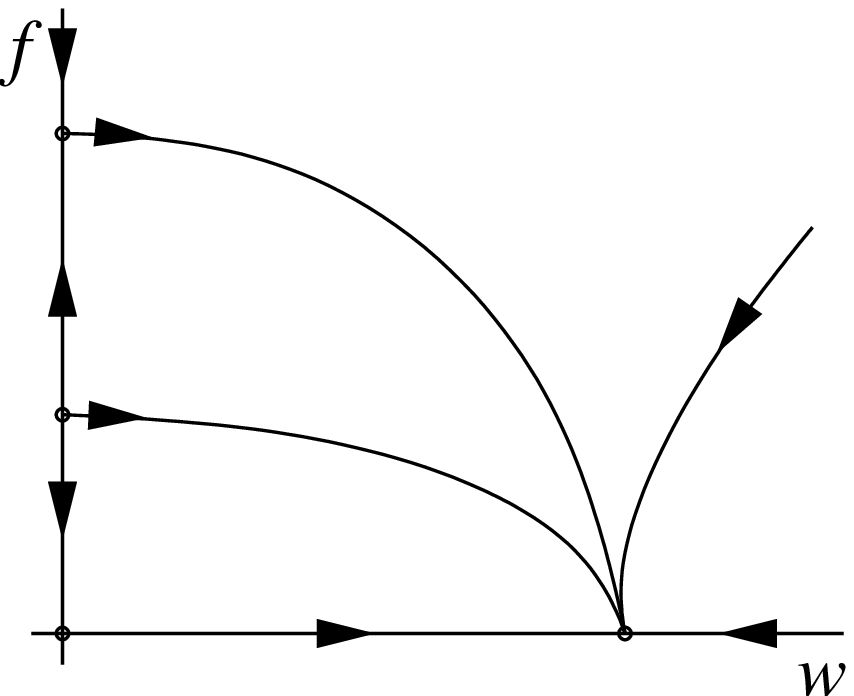,width=0.35\linewidth}
\epsfig{file=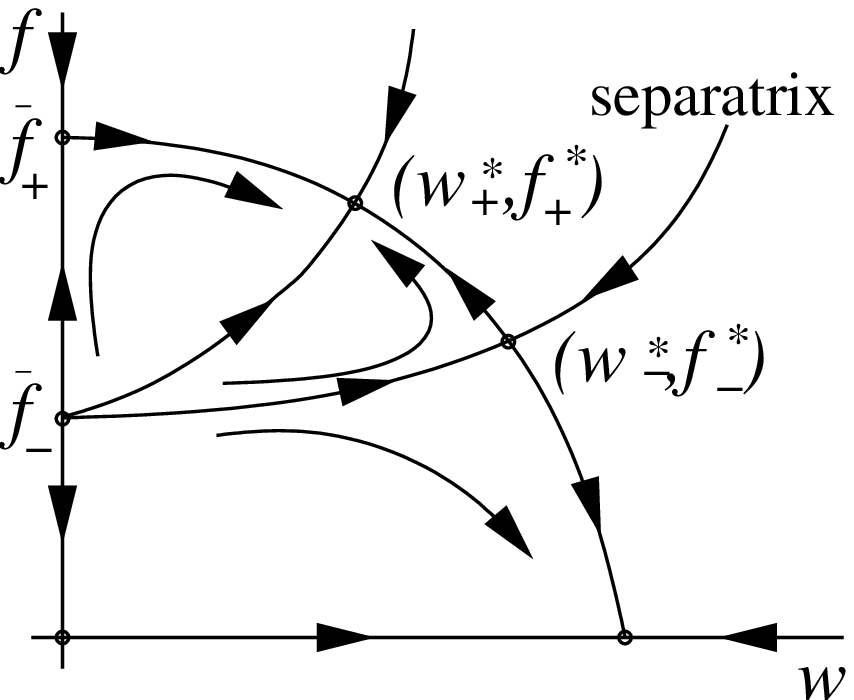,width=0.35\linewidth}
\end{center}

\caption{Two of the possible phase planes for (\ref{eq:fdp}),
(\ref{eq:wdp}) illustrating the saddle--node bifurcation. Left:
$\dam<\dam_c^{\rm pulse}$, and the fixed points $P_4$ do not
exist. Right: $\dam>\dam_c^{\rm pulse}$.}

\label{fig:pp}
\end{figure}

Two phase planes for (\ref{eq:fdp}), (\ref{eq:wdp}) illustrating the
saddle--node bifurcation at $\dam_c$ are shown in Figure~\ref{fig:pp}. If
$\dam<\dam_c^{\rm pulse}$, the fixed points $P_4$ do not exist, and all
initial conditions off the $f$-axis asymptote to the solution
$(\steady{w},0)$ at large time (physically the trivial state $u\equiv0$).
If instead $\dam>\dam_c^{\rm pulse}$, the $P_4$ solutions exist, and there
is a separatrix that divides initial conditions leading to the physically
trivial state $(\steady{w},0)$ from those leading to the physically
nontrivial state $(w^*_+,f^*_+)$. 

\begin{figure}
\begin{center}
\epsfig{file=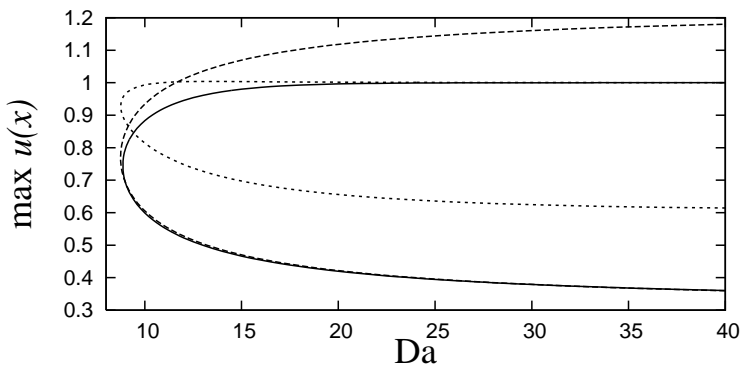,width=0.6\linewidth}
\end{center}

\caption{Comparison between the steady states of (\ref{eq:ard}) computed
by shooting (solid line), the pulse model (\ref{eq:P4}) (dashed line)
and the plateau model (\ref{eq:ABC}) (dotted line). Here $\alpha=0.2$.}

\label{fig:pulse}
\end{figure}

\begin{figure}
\begin{center}
\epsfig{file=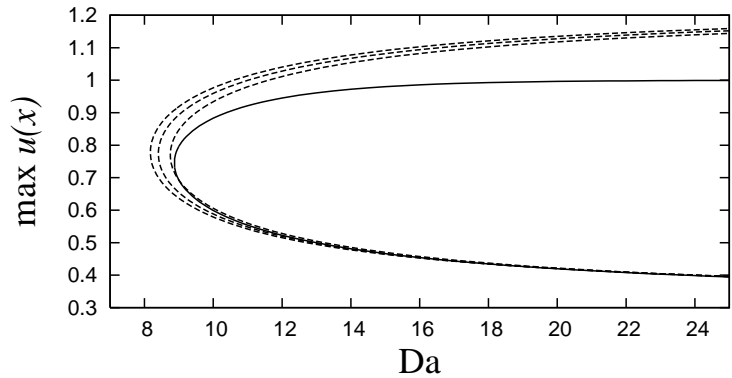,width=0.6\linewidth}
\end{center}

\caption{Comparison between different shaped pulses in the model
(\ref{eq:P4}). The solid curve represents the steady states of
(\ref{eq:ard}) computed by shooting; the dashed lines represent
solutions of the pulse model (\ref{eq:P4}) for test functions
$U(\eta)=\e^{-\eta^2}$, $\sech^2\eta$ and $\sech^4\eta$ (respectively
from right to left at the saddle--node bifurcation point). 
Parameters as in Figure~\ref{fig:pulse}.}

\label{fig:pulse-sech2}
\end{figure}

Figure~\ref{fig:pulse} shows a comparison between results from the
Gaussian pulse model and steady-state solutions of (\ref{eq:ard})
computed using shooting. The variational result (\ref{eq:P4})
accurately reproduces the bifurcation behaviour, both in terms of the
critical Damk\"ohler number and the generic quadratic behaviour close
to the saddle--node bifurcation. A motivation for our Gaussian
ansatz was given by the small-$\dam$ asymptotic argument in
Section~\ref{Sec_asympt2}. However, in Figure~\ref{fig:pulse-sech2} we show
that good agreement is also given for different ans{\"a}tze such as
$U(\eta)=\sech^n\eta$. We thus conclude that the actual form of the pulse
shape is not particularly important (and hence we cannot expect to
deduce this pulse shape from any asymptotic theory).

\begin{figure}
\begin{center}
\epsfig{file=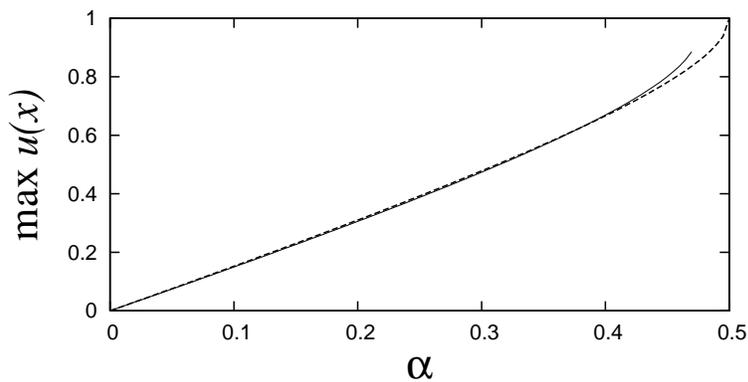,width=0.6\linewidth}
\end{center}

\caption{Asymptotic value of $\max u(x)$ for the unstable steady state at
large $\dam$. Solid line: result from (\ref{eq:P4}). Dashed line:
result from (\ref{eq:V(0)}).}

\label{fig:pmax}
\end{figure}

By contrast, the asymptotic approaches which are
usually employed, and which assume plateau-like solutions with
well-separated constant plateaus, fail to describe the bifurcation
behaviour. We note that the agreement between the pulse model and the full
problem (\ref{eq:ard}) is excellent on the lower branch.
Figure~\ref{fig:pmax} shows the excellent agreement between $u(0)=\max
u(x)$ computed according to the pulse result (\ref{eq:P4}) in the limit
$\dam\to\infty$ and a direct large-$\dam$ asymptotic analysis of
(\ref{eq:ard}), given by (\ref{eq:V(0)}). 

\section{Plateau solutions}
\label{Sec_fronts}

Away from the saddle--node bifurcation, the stable solution takes
the form of a nearly uniform region around $x=0$, surrounded by a
pair of fronts. This solution may be captured by writing $u(x,t)$ in the form
\beq
u(x,t)=\tsfrac12 f(t)\phi(\eta) \qquad\mbox{with}\qquad \eta=w(t)x,
\label{eq:An}
\eeq
where
\[
\phi(\eta)=\tanh(\eta+a(t))-\tanh(\eta-a(t)).
\]
Then the evolution of $f(t)$, $w(t)$ and $\nu(t)=a/w$ (or, equivalently,
$f$, $w$ and $a$) may be determined by forcing the inner products of 
(\ref{eq:ard}) with each of $u_f$, $u_w$ and $u_\nu$ to be zero, where
\bean
u_f&=&\pder{u}{f}=\tsfrac12\phi(\eta)=\frac{u}{f}\\
u_w&=&\pder{u}{w}=
\tsfrac12f\left[(x+\nu){\sech}^2(\eta+a)-(x-\nu)\sech^2(\eta-a)\right]\\
u_\nu&=&\pder{u}{\nu}=\tsfrac12 fw\psi(\eta),
\eean
with
\[
\psi(\eta)=\sech^2(\eta+a)+\sech^2(\eta-a).
\]

The corresponding steady-state calculation has been considered in
\cite{Menon05}; here we extend their analysis to the time-dependent
problem. This permits us to determine which initial conditions lead
to excitation and which to extinction. Details of our derivation,
including analytical expressions for
all the requisite inner products, are given in the Appendix: we find that
the equations for the time evolution of $f$, $w$ and $\nu$ take the form
\bea
\lambda_1 w\dot{f}+\lambda_2 f\dot{w}+\lambda_3w^2f\dot{\nu}&=&wf\left[
\lambda_4+\lambda_5Dw^2+\dam(\lambda_6+\lambda_7f+\lambda_8f^2)\right]
\label{eq:td1}\\
\mu_1 w\dot{f}+\mu_2 f\dot{w}+\mu_3w^2f\dot{\nu}&=&wf\left[
\mu_4+\mu_5Dw^2+\dam(\mu_6+\mu_7f+\mu_8f^2)\right]
\label{eq:td2}\\
\beta_1 w\dot{f}+\beta_2 f\dot{w}+\beta_3w^2f\dot{\nu}&=&wf\left[
\beta_4+\beta_5Dw^2+\dam(\beta_6+\beta_7f+\beta_8f^2)\right].
\label{eq:td3}
\eea
The various coefficients $\lambda_i$, $\mu_i$ and $\beta_i$ are
complicated functions of $a$ and $\alpha$, and are listed in the Appendix.

\subsection{Steady states}

We shall not attempt here a full analysis of the three-dimensional phase space
for $f$, $w$ and $\nu$; instead we consider only nontrivial steady-state
solutions for which $f$, $w$ and $\nu$ are all nonzero. We expect two such
solutions when $\dam$ exceeds some threshold, one stable and one unstable.
We shall consider the dynamics of (\ref{eq:td1})--(\ref{eq:td3}) later.

The equations for the (nontrivial) steady states may be written in the
form
\bea
\lambda_4+\lambda_5Dw^2+
\dam(\lambda_6+\lambda_7f+\lambda_8f^2)&=&0\label{eq:ss1}
\\
\mu_4+\mu_5Dw^2+\dam(\mu_6+\mu_7f+\mu_8f^2)&=&0\label{eq:ss2}\\
\beta_4+\beta_5Dw^2+\dam(\beta_6+\beta_7f+\beta_8f^2)&=&0.\label{eq:ss3}
\eea
These equations are readily solved without recourse to nonlinear root finding,
for example in the following way. First we fix values for $\alpha$ and $a$.
Then we use (\ref{eq:ss1}) to write
\beq
w^2=\frac{-\lambda_4-\dam(\lambda_6+\lambda_7f+\lambda_8f^2)}{\lambda_5D}.
\label{eq:w2}
\eeq
Then, upon substitution of this expression for $w^2$ in (\ref{eq:ss2}), we
have
\beq
\mu_4'+\dam(\mu_6'+\mu_7'f+\mu_8'f^2)=0,\label{eq:mu'} \quad
{\rm{where}} \quad 
\mu_i'=\mu_i-\frac{\mu_5}{\lambda_5}\lambda_i.
\eeq
Likewise
\beq
\beta_4'+\dam(\beta_6'+\beta_7'f+\beta_8'f^2)=0,\label{eq:beta'} \quad
{\rm{where}} \quad 
\beta_i'=\beta_i-\frac{\beta_5}{\lambda_5}\lambda_i.
\eeq
Now by eliminating $\dam$ between (\ref{eq:mu'}) and (\ref{eq:beta'}),
we find the following equation for $f$:
\beq
(\mu_4'\beta_8'-\mu_8'\beta_4')f^2+(\mu_4'\beta_7'-\mu_7'\beta_4')f
+\mu_4'\beta_6'-\mu_6'\beta_4'=0,
\label{eq:fq}
\eeq
which is of the form
\beq
A(a)f^2+(1+\alpha)B(a)f+\alpha C(a)=0.
\label{eq:ABC}
\eeq

\begin{figure}

\begin{center}
\epsfig{file=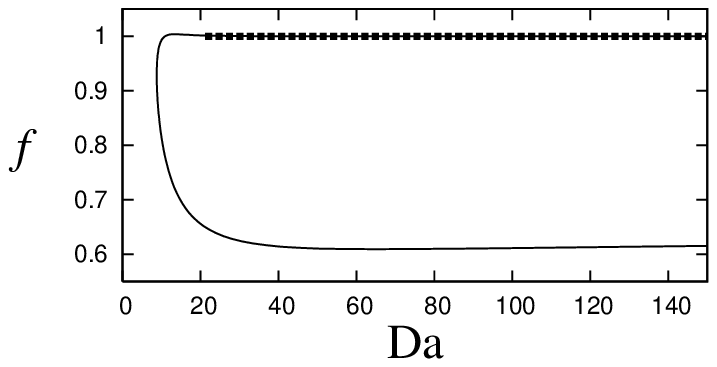,width=0.4\linewidth}
\epsfig{file=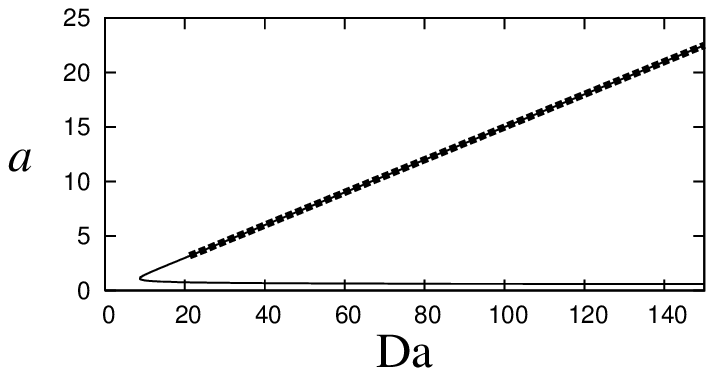,width=0.4\linewidth}

\epsfig{file=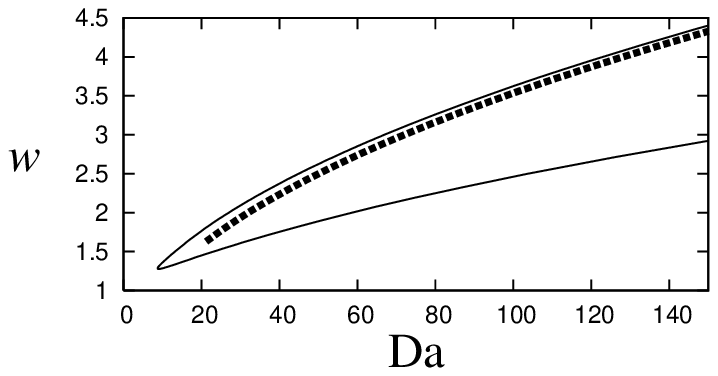,width=0.4\linewidth}
\epsfig{file=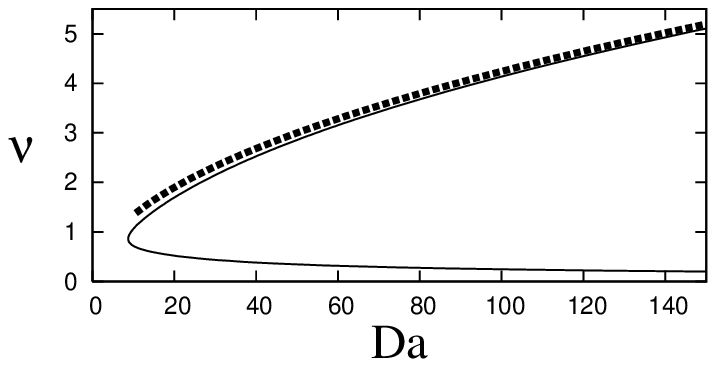,width=0.4\linewidth}

\end{center}

\caption{Steady-state solutions of the plateau model
(\ref{eq:ss1})--(\ref{eq:ss3}) (solid line) for $\alpha=0.2$. The dashed
lines give large-$\dam$ asymptotics (\ref{eq:fwa}), (\ref{nu}) of the
plateau model for the upper branch. The asymptotic results are
indistinguishable from the numerical results for $f$ and $a$.}

\label{fig:steady0} 
\end{figure}

\begin{figure}
\begin{center}
\epsfig{file=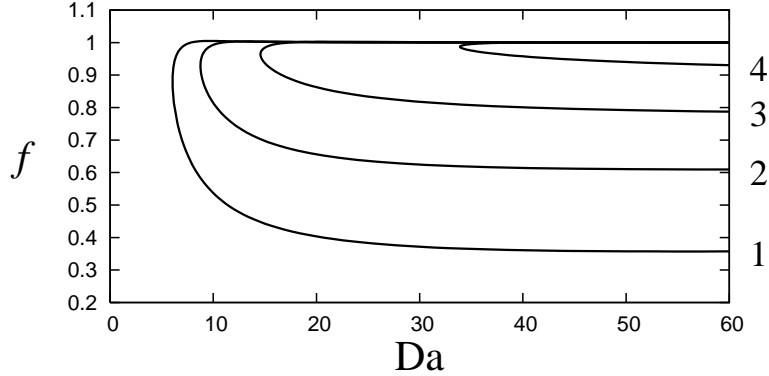,width=0.6\linewidth}
\end{center}

\caption{Steady-state solutions of the plateau model
(\ref{eq:ss1})--(\ref{eq:ss3}) for $\alpha=0.1,0.2,0.3,0.4$. (The 
corresponding value of $10\alpha$ is indicated next to each curve.)}

\label{fig:steady1}
\end{figure}

The quadratic equation (\ref{eq:ABC}) needs careful interpretation.
First we note that the coefficients $A$, $B$ and $C$ are functions of
$a$ alone, and hence for given values of $a$ and $\alpha$ there are in
general two solutions of (\ref{eq:fq}) or none. When there are two
solutions, they correspond, through (\ref{eq:mu'}) or
(\ref{eq:beta'}), to {\em distinct} values of $\dam$; numerically, we
observe that one value of $\dam$ is positive and the other negative
(and hence physically unreasonable).  Furthermore, when we reconstruct
from (\ref{eq:w2}) the corresponding values of $w$, we find
numerically that the root of (\ref{eq:fq}) corresponding to positive
$\dam$ seems always to give rise to $w^2>0$, and so is indeed
physically reasonable.  Thus we emphasise that both branches of
solutions evident in Figures~\ref{fig:steady0} and~\ref{fig:steady1}
arise from {\em one} of the roots of (\ref{eq:ABC}) for fixed
$\alpha$, as $a$ is increased monotonically.

The steady states for $\alpha=0.2$ (see \cite{Menon05}) are shown in
Figure~\ref{fig:steady0}. Note that solutions exist only for
$\dam>\dam_c^{\rm plateau}(\alpha)$. Figure~\ref{fig:steady1} shows
steady-state solution branches for a range of values of $\alpha$.  The
excellent agreement between solutions of the plateau model and steady
states of (\ref{eq:ard}) on the upper branch is illustrated in
Figures~\ref{fig:pulse} and~\ref{fig:an}. Of course, we expect a
$\tanh$-test function ansatz to fail close to the bifurcation point
$\dam_c$ where the solution rather is bell-shaped and is accurately
described by a Gaussian ansatz as in Section \ref{Sec_pulses}.

Of particular interest is the limit of large Damk\"ohler number,
$\dam\gg1$.  
The numerics suggest the following scalings.  On the upper branch of
solutions
\beq
a=\bar{a}\dam,\qquad w=\bar{w} \dam^{1/2},\qquad f=\bar{f},
\label{eq:scalu}
\eeq
where $\bar{a},\bar{w},\bar{f}=\Ord(1)$; on the lower branch
\[
a=\tilde{a},\qquad w=\tilde{w} \dam^{1/2},\qquad f=\tilde{f},
\]
where $\tilde{a},\tilde{w},\tilde{f}=\Ord(1)$.

\subsubsection{Large Damk\"ohler number, $\dam\gg1$, upper branch}
\label{sec:5.1.1}

We now construct the plateau solutions on the upper branch, in the limit
$\dam\gg1$. In that limit, we find
\[
\begin{array}{lclcl}
\lambda_4\sim -a  & &
\mu_4\sim\tsfrac{1}{18}\pi^2-\tsfrac13  & &
\beta_4\sim-\tsfrac23a  \\
\lambda_5\sim -\tsfrac23  & &
\mu_5\sim-\tsfrac13  & &
\beta_5\to0  \\
\lambda_6\sim -2a\alpha  & &
\mu_6\sim-\tsfrac12\alpha  & &
\beta_6\sim-\alpha  \\
\lambda_7\sim2(1+\alpha)a  & &
\mu_7\sim\tsfrac12(1+\alpha)  & &
\beta_7\sim\tsfrac23(1+\alpha)  \\
\lambda_8\sim-2a  & &
\mu_8\sim-\tsfrac{11}{24}  & &
\beta_8\sim-\tsfrac12.
\end{array}
\]
Recall that $a=\Ord(\dam)$. The result for $\beta_5$ indicates
that this quantity is exponentially small in $\dam$.  Thus at leading
order in $\dam$ we have, from (\ref{eq:ss1})--(\ref{eq:ss3}),
\bean
-\alpha+(1+\alpha)\bar{f}-{\bar{f}}^2&=&0\\
-\tsfrac13D{\bar{w}}^2+\tsfrac12(-\alpha+(1+\alpha)\bar{f}-\tsfrac{11}{12}
{\bar{f}}^2)&=&0\\
-\tsfrac23\bar{a}+(-\alpha+\tsfrac23(1+\alpha)\bar{f}-\tsfrac12{\bar{f}}^2)
&=&0.
\eean
The first of these equations gives $\bar{f}=1$ or $\bar{f}=\alpha$; the latter
solution must be rejected since the last equation would then give
$\bar{a}=-(2-\alpha)\alpha/4<0$. The choice $\bar{f}=1$ is consistent
with our numerical calculations, which indicate that this is the solution
relevant to the upper branch. With $\bar{f}=1$, the remaining equations
give $D{\bar{w}}^2=\tsfrac18$ and $\bar{a}=\tsfrac14(1-2\alpha)$. Thus in the
large-$\dam$ limit,
\beq
f\sim1,\qquad
w\sim\left(\frac{\dam}{8D}\right)^{1/2},\qquad
a\sim\tsfrac14(1-2\alpha)\dam.
\label{eq:fwa}
\eeq
Note, as a consequence, that for large $\dam$ we have
\beq
\nu\sim\sqrt{2D\dam}\left(\tsfrac12-\alpha\right),
\label{nu}
\eeq
in agreement with the phenomenological argument given in
\cite{Neufeld02,Menon05}. This formula can be understood if we note that
the location of a stationary front is given approximately through a
balance between the front velocity
$v_0=\sqrt{2D\dam}\left(\frac{1}{2}-\alpha\right)$
of the unstirred problem \cite{Neufeld02} and the velocity of the
chaotic stirring (i.e., $x$ in the scaling of (\ref{eq:ard})). Thus in
the stirred problem, the front has zero velocity when $v_0=x$, which
implies $v_0=\nu$, and (\ref{nu}) follows.

\begin{figure}
\begin{center}
\epsfig{file=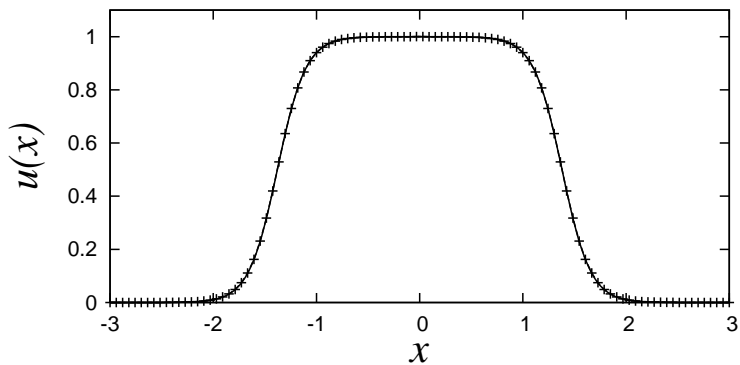,width=0.6\linewidth}
\end{center}

\caption{Comparison between stable steady states of the plateau model
(crosses) and of the advection--reaction--diffusion equation
(\ref{eq:ard}) (solid line). Here $\alpha=0.4$ and $\dam=100.0359$
(implied by our choice of $a$); for the plateau-model solution, $a=5.0$,
$f=1.000179$ and $w=3.62460$.}

\label{fig:an}
\end{figure}

Finally, the solution takes the form
\beq
u\sim\frac12\left\{
\tanh\left[\sqrt{\frac{\dam}{8D}}x+\frac{(1-2\alpha)\dam}{4}\right]-
\tanh\left[\sqrt{\frac{\dam}{8D}}x-\frac{(1-2\alpha)\dam}{4}\right]
\right\}.
\eeq
A comparison between this expression and a numerical calculation
of the stable steady state of (\ref{eq:ard}) is shown in Figure~\ref{fig:an};
to graphical resolution, the two graphs are indistinguishable.

\begin{figure}
\begin{center}
\epsfig{file=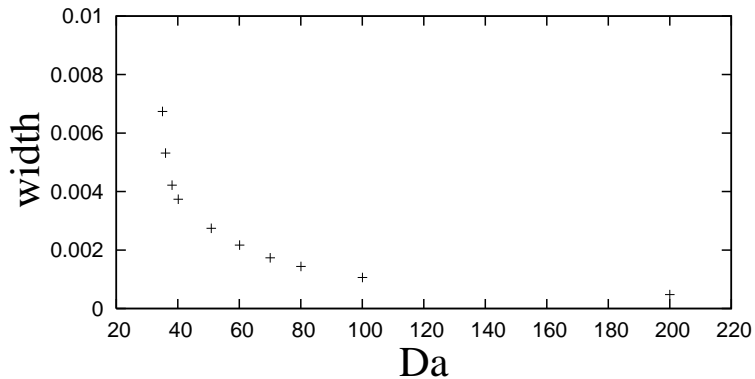,width=0.6\linewidth}
\end{center}

\caption{Comparison between the width of the steady-state solution as
obtained from the plateau model and from the advection--reaction--diffusion
equation (\ref{eq:ard}). Here $\alpha=0.4$. Shown is $2(\nu_{\rm
approx}-\nu_{\rm exact})$ against the Damk\"ohler number $\dam$.}

\label{fig:width}
\end{figure}

Figure~\ref{fig:width} shows how well the ansatz (\ref{eq:An})
captures the width of the steady-state solution. Plotted is
$2(\nu_{\rm approx}-\nu_{\rm exact})$, where $\nu_{\rm approx}$ and
$\nu_{\rm exact}$ are, respectively, the half-widths of the solutions
to the plateau model and the full problem (\ref{eq:ard}). In each case
only the stable solution on the upper branch is considered, and the
`half-width' is defined as $\nu_{\rm approx}, \nu_{\rm exact}=x_+>0$,
where $u(x_+)=0.5$. The difference $2(\nu_{\rm approx}-\nu_{\rm
exact})$ clearly decreases with increasing Damk\"ohler number, which
is expected, because the $\tanh$-front ansatz gets closer to the
actual solution, which asymptotically is a pair of $\tanh$-fronts (see
Section~\ref{Sec_asympt}).

Solutions of the plateau model on the lower branch are of less interest
because they do not correspond closely to steady-state solutions of the
full PDE, where the unstable solution is pulse-like. 

\subsection{Stability}

Suppose that $(f,w,\nu)$ is a nontrivial steady state of
(\ref{eq:td1})--(\ref{eq:td3}) (i.e., $f,w,\nu\neq0$). Since the various
coefficients in (\ref{eq:td1})--(\ref{eq:td3}) are functions of $a$, it is
useful to consider the linearised evolution of $\delta f$, $\delta w$ and
$\delta a$, where in the linearisation $\delta a=w\delta\nu+ \nu\delta w$.
Thus we have
\bean
\lambda_1 w\delta \dot{f}+\lambda_2 f\delta \dot{w}+\lambda_3 f w 
(\delta\dot{a}-\nu\delta\dot{w})&=&
\dam\, fw(\lambda_7+2\lambda_8 f)\delta f
+2Dfw^2\lambda_5\delta w\\
&&{}+fw\left\{
\partial\lambda_4+Dw^2\partial\lambda_5+\dam(\partial\lambda_6
+f\partial\lambda_7+f^2\partial\lambda_8) \right\}\delta a,
\eean
where
\[
\partial\lambda_i\equiv\pder{\lambda_i}{a}
\]
and where all nonperturbation quantities are evaluated at the fixed point.
There are two further equations, with $\lambda_i\mapsto\mu_i$ and
$\lambda_i\mapsto\beta_i$. The three may be written together as
\[
{\mathcal A}\left(\begin{array}{c}\delta\dot{f}\\ 
\delta\dot{w}\\
\delta\dot{a}\end{array}\right)=
{\mathcal B}
\left(\begin{array}{c}\delta f\\ \delta w\\ \delta a\end{array}\right),
\]
where the constant-coefficient matrices ${\mathcal A}$ and ${\mathcal B}$ 
are readily determined from the three perturbation evolution equations.

\subsubsection{Large Damk\"ohler number, $\dam\gg1$, upper branch}

On the upper solution branch, in the limit $\dam\gg1$, we have
\[
{\mathcal A}\sim\left(\begin{array}{rrr}
2aw & -a & w \\
\tsfrac12 w & \tsfrac{1}{18}(\pi^2-6) & 0 \\
w & -\tsfrac23 a & \tsfrac23w 
\end{array}\right),\qquad
{\mathcal B}\sim\left(\begin{array}{rrr}
-2\dam\, wa(1-\alpha) & -\tsfrac43Dw^2 & -w \\
\tsfrac1{12}\dam \,w(6\alpha-5) & -\tsfrac23Dw^2 & 0\\
-\tsfrac13\dam\, w(1-2\alpha) & 0 & -\tsfrac23w
\end{array}\right),
\]
where the zero entries indicate terms that are exponentially small in $\dam$.
Thus, in view of the scalings (\ref{eq:scalu}),
\[
{\mathcal A}\sim\left(\begin{array}{rrr}
2\bar{a}\bar{w}\dam^{3/2} & -\bar{a}\dam & \bar{w}\dam^{1/2} \\
\tsfrac12 \bar{w}\dam^{1/2} & \tsfrac{1}{18}(\pi^2-6) & 0\\
\bar{w}\dam^{1/2} & -\tsfrac23 \bar{a}\dam & \tsfrac23\bar{w}\dam^{1/2}
\end{array}\right)
\]
and
\[
{\mathcal B}\sim\left(\begin{array}{rrr} 
-2\bar{w}\bar{a}(1-\alpha)\dam^{5/2} &
-\tsfrac43D\bar{w}^2\dam & -\bar{w}\dam^{1/2} \\
\tsfrac1{12}\bar{w}(6\alpha-5)\dam^{3/2} & -\tsfrac23D\bar{w}^2\dam & 0\\
-\tsfrac13\bar{w}(1-2\alpha)\dam^{3/2} & 0 & -\tsfrac23\bar{w}\dam^{1/2}
\end{array}\right).
\]
It then follows (after some algebra) that\
\[
{\mathcal A}^{-1}{\mathcal B}\sim\left(\begin{array}{rrr}
-(1-\alpha)\dam & -\dsfrac{2D\bar{w}}{3\bar{a}}\dam^{-1/2} &  0 \\
\dsfrac{3\bar{w}}{2(\pi^2-6)}\dam^{3/2} &
-\dsfrac{3}{2(\pi^2-6)}\dam & 0\\
\dsfrac{3\bar{a}}{2(\pi^2-6)}\dam^2 &
-\dsfrac{12D\bar{w}\bar{a}}{\pi^2-6}\dam^{3/2} & -1 
\end{array}\right).
\]

We require the dominant eigenmodes of ${\mathcal A}^{-1}{\mathcal B}$. There 
are three eigenvalues:
\beq
\Lambda_1\sim-1\qquad\mbox{and}\qquad \Lambda_2,\Lambda_3=\Ord(\dam).
\label{eq:Lambdas}
\eeq
The eigenvector corresponding to $\Lambda_1$ has $\delta f=\delta w=0$ and
$\delta a=1$, say; thus disturbances in $a$ decay like $\e^{-t}$. We find
$\Lambda_2\sim-(1-\alpha)\dam$; recalling the scalings $f=\Ord(1)$,
$w=\Ord(\sqrt{\dam})$ and $\nu=\Ord(\sqrt{\dam})$, we find that the
associated eigenvector has $(\delta f,\delta w,\delta
a)\sim(F,W\dam^{1/2},A\dam)$, with
\[
\left(-(1-\alpha)+\frac{3}{2(\pi^2-6)}\right)W=
\frac{3\bar{w}}{2(\pi^2-6)}F 
\quad {\rm{and}} \quad
-(1-\alpha)A=\frac{3\bar{a}}{2(\pi^2-6)}F-\frac{12D\bar{w}\bar{a}}
{\pi^2-6}W.
\]
The final eigenvalue is 
\[
\Lambda_3\sim-\frac{3}{2(\pi^2-6)}\dam,
\]
with eigenvector $(\delta f,\delta w,\delta a)\sim(F,W\dam^{3/2},A\dam^2)$,
where
\[
\left(-\frac{3}{2(\pi^2-6)}+1-\alpha\right)F=
-\frac{2D\bar{w}}{3\bar{a}}W 
\quad {\rm{and}} \quad
A=8D\bar{w}\bar{a}W.
\]
Note that this calculation confirms the stability of the upper branch solution,
since $\Lambda_{1,2,3}<0$.  Furthermore, the scalings (\ref{eq:Lambdas})
confirm the observation from numerical simulations of the PDE (\ref{eq:ard})
that the relaxation of the front separation to its equilibrium value
is slower than the relaxation of either the amplitude of the solution or
the width of each front.
(Provided the fronts are reasonably well separated initially, numerical
integrations of either the PDE or the system (\ref{eq:td1})--(\ref{eq:td3}) 
indicate that first $f$ adjusts to approximately its long-time value;
then $w$ adjusts to give each front the appropriate width, and finally
$a$ adjusts so that the fronts are separated by the correct distance.)

Since $0<\alpha<\tsfrac12$ and since $\tsfrac32(\pi^2-6)^{-1}\approx
0.3876$, we have
\[
|\Lambda_2|>|\Lambda_3|\gg|\Lambda_1|.
\]


\section{Comparison between dynamics of the reduced models and the
full PDE}
\label{Sec_initial}

\begin{figure}
\begin{center}
\epsfig{file=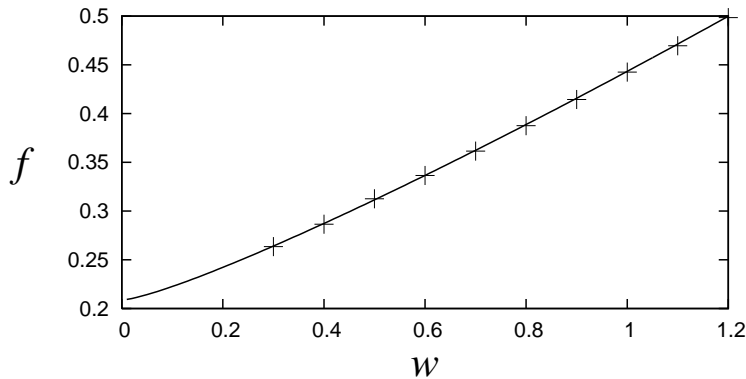,width=0.6\linewidth}
\end{center}

\caption{Comparison between the separatrix $f=f_{s}^{\rm pulse}(w)$ determined
from the Gaussian pulse ansatz (continuous line) and the threshold
$f=f_{s}^{\rm PDE}(w)$ obtained from simulations of the full PDE
(\ref{eq:ard}) (crosses) with initial profile $u(x,0)=f\e^{-(wx)^2}$
as a function of $w$; parameter values are $\alpha=0.2$ and $\dam=16$.}

\label{fig:table1plot}
\end{figure}

Our models extend the steady-state analysis of \cite{Menon05} by
providing approximations to the {\em dynamics} of the PDE
(\ref{eq:ard}). This allows us to examine the evolution of initial
conditions which are either plateau-like or pulse-like. For instance,
using the Gaussian pulse ansatz, the dynamics is approximately reduced
to that of the two-dimensional system (\ref{eq:fdp})--(\ref{eq:wdp}).
Given an initial pulse-like solution we may now solve the
initial-value problem at Damk\"ohler numbers close to the critical
Damk\"ohler number $\dam_c$, where both stable and unstable solutions
are well approximated by a bell-shaped function. We can readily
compute the corresponding separatrix in $(w,f)$-space between those
solutions that asymptote at large times to the uniform zero solution,
and those that asymptote to a nontrivial steady state. We denote the
separatrix by $f=f_s^{\rm pulse}(w)$. For initial conditions
$(w(0),f(0))$ below the separatrix the solution will die out, whereas
for initial conditions above the separatrix the reaction will continue
and develop into a stable stationary pulse. It is of particular
interest to determine the extent to which this result provides useful
corresponding information for the full PDE (\ref{eq:ard}). We have
therefore carried out a large number of simulations of the PDE, each
from a Gaussian initial condition $u(x,0)=f\e^{-(wx)^2}$, in an effort
to compute a corresponding separatrix: $f=f_s^{\rm PDE}(w)$.  For a
given value of $w$, we use bisection to determine approximately the
corresponding value of $f$ that delineates the two large-time
behaviours. The results, and a comparison with the separatrix from the
Gaussian pulse ansatz, are given in Figure~\ref{fig:table1plot}. The
agreement is excellent.  There is similar agreement at higher $\dam$;
for example, at $\dam=100$, and $w(0)=1$, the Gaussian ansatz gives a
threshold $f(0)=0.2537$, while from the PDE we find $0.2525\pm0.0005$.





\begin{figure}
\begin{center}
\epsfig{file=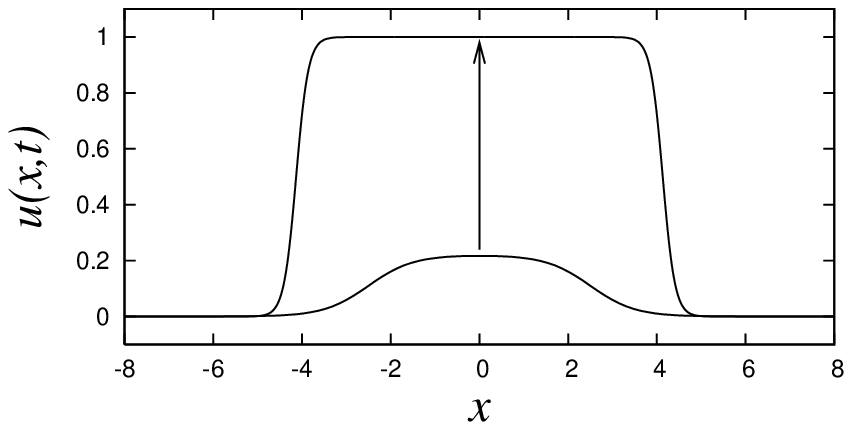,width=0.5\linewidth}

\epsfig{file=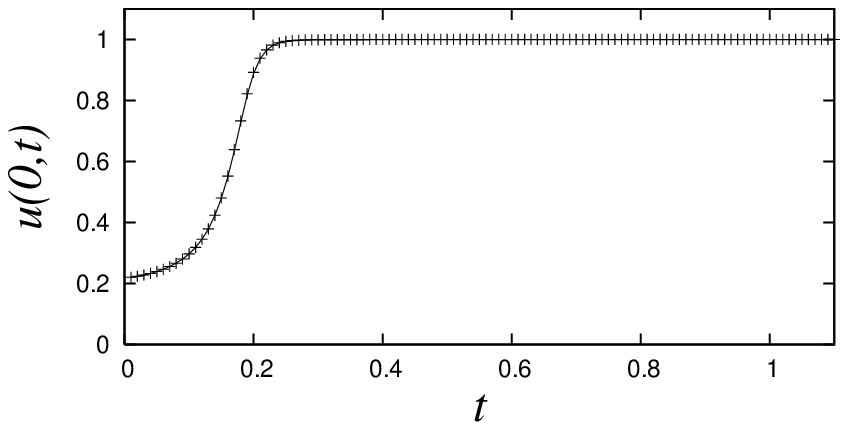,width=0.4\linewidth}
\epsfig{file=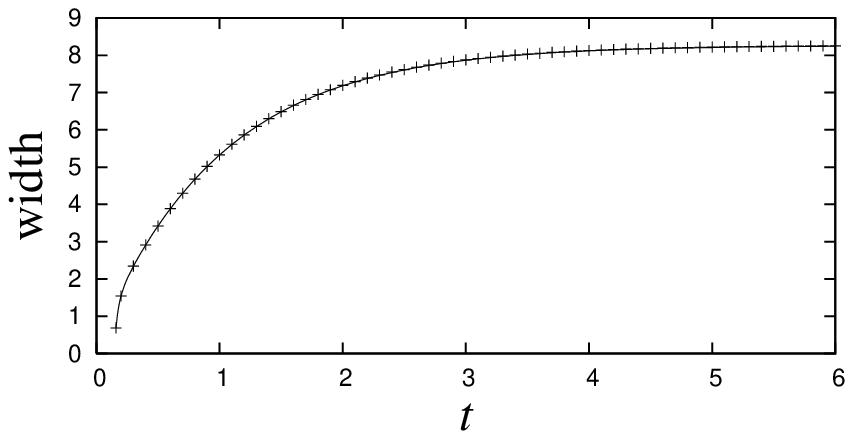,width=0.4\linewidth}
\end{center}

\caption{Comparison between numerical simulations of the PDE (\ref{eq:ard})
and the plateau model (\ref{eq:td1})--(\ref{eq:td3}).
Top: initial and final states, with arrow indicating direction of temporal 
evolution.  Bottom left: $u(0,t)$ against time. Bottom right: `width' of the
pulse, $2\nu$.  In each of the lower plots, PDE results are given by
a solid line and plateau-model results by crosses.}

\label{fig:testf}
\end{figure}

Assessing the usefulness of the plateau model is rather harder. This is
because if we start with an initial profile of the form (\ref{eq:An}),
with well separated fronts, and $u(x,0)$ approximately uniform between those
fronts, then in the uniform region the initial evolution of $u$ is well
approximated by the ODE
\[
\deriv{u}{t}=\dam\, u(\alpha-u)(u-1),
\]
so that $f$ is initially decreasing if $f(0)<\alpha$ and increasing if
$f(0)>\alpha$. Thus the threshold value of $f(0)$ leading to zero or nonzero
large-time states is rather trivially $f(0)\approx\alpha$.

The agreement between numerical integrations of the PDE (\ref{eq:ard}) and
the ODE system for a plateau-like solution (\ref{eq:td1})--(\ref{eq:td3}) is
illustrated in Figure~\ref{fig:testf}, which shows simulations from an
initial plateau-like state of the form (\ref{eq:An}), for $\alpha=0.2$ and
$\dam=100$. The initial condition is $f=0.22$, $w=1$ and $a=2.5$. Two
quantities are plotted: the maximum $u(0,t)$ and the `width' of the
solution, $2\nu$, defined again as $2x_+$, where $u(x_+,t)=0.5$ and
$x_+>0$. (Note that initially the entire solution lies below $0.5$,
so the `width' is initially undefined.) The solid
lines represent numerical simulations of the PDE (\ref{eq:ard}); the
crosses represent simulations of the ODE system
(\ref{eq:td1})--(\ref{eq:td3}). The agreement is excellent, indicating
that at large $\dam$ the plateau model accurately describes not only the
stable steady state, but also the dynamics of solutions to the PDE. 

The agreement between the plateau model and the full PDE is remarkable
because, as shown in Section~\ref{Sec_asympt}, the unstable solution is
pulse-like and not plateau-like.  Furthermore, as shown in
Figure~\ref{fig:pulse}, the unstable solution predicted by the plateau model
is quantitatively incorrect. However, some insight into the agreement may
be obtained by noting that the unstable pulse-like steady-state has
eigenvalues $\sigma=\Ord(\dam)$, which are large in the limit
$\dam\to\infty$, and which indicate that the associated instability is
reaction-dominated.  Our simulations of the PDE (\ref{eq:ard}) confirm
that the initial condition illustrated in Figure~\ref{fig:testf} first
rapidly undergoes an adjustment of the central concentration to
$u\approx1$, consistent with a time scale $\Ord(1/\dam)$, then a slower
relaxation of the front width and location. 


\section{Conclusions}

We have developed reduced systems of ordinary differential equations
describing the time evolution of both plateau-like and pulse-like solutions
of a one-dimensional bistable reaction--diffusion partial differential
equation for a stirred flow environment. The PDE possesses a trivial
steady state in which all the reactant is used up, and, when the
dimensionless reaction rate $\dam$ is great enough, a pair of nontrivial
steady states, one stable and one unstable, born in a saddle--node bifurcation
at $\dam=\dam_c$. The pulse model was shown to provide a good approximation
close to the bifurcation point and for the unstable steady solution at
large $\dam$. The plateau model applies for large $\dam$, and captures well the
form of the stable steady state.  By studying equilibrium pulse solutions,
we were able to accurately describe the bifurcation behaviour of the
system near $\dam_c$. For large Damk\"ohler numbers our reduced plateau
model allowed us to describe the stationary fronts, and derive a
commonly used phenomenological formula for the width of a stationary front. 

The time-dependent ODE models allowed us to study the initial-value
problem for (\ref{eq:ard}). We found that although the unstable
solution is actually of a pulse-like shape, the dynamics of (\ref{eq:ard})
is well captured by the plateau model for high Damk\"ohler numbers. 
This remarkable and nonintuitive result is due to the fact that the
growth rate of the instability of the unstable solution is
$\Ord(\dam)$, whereas the fastest growth rate for the shape-change of
a front is $\Ord(1)$. Our numerical solutions of (\ref{eq:ard}) seem
rapidly to approach the plateau form, and thus the dynamics of the PDE
(\ref{eq:ard}) is described accurately by the plateau model.

Besides being instructive in themselves, one-dimensional models such as 
(\ref{gen_eqn}) aim to provide information about the evolution of 
chemical reactions in a two- or three-dimensional chaotic flow.
It is now known \cite{Cox04} that simple variants of reduced models 
perform poorly in predicting the yield of multi-stage reactions, and so
a more sophisticated analysis, as presented here for the bistable case, is 
therefore warranted. One particularly promising extension \cite{Vikhansky04}
of one-dimensional models such as (\ref{gen_eqn}) 
involves two stages. In the first stage, a point $\vec x$ in the
two- or three-dimensional problem and a time $t$ are selected,
at which the chemical composition is required.  This point and its
associated stable manifold are then advected backwards in time,
to $t=0$, and the initial condition for the system is imagined
to be sampled by the stable manifold (which is now exponentially
stretched and highly contorted). The second stage involves carrying
out a one-dimensional simulation of the advection, reaction and diffusion
along the evolving stable manifold (now forwards in time), using
a model such as (\ref{gen_eqn}), but with a nonconstant compression
rate~\cite{CV06}.

\clearpage

\section*{Appendix}

Here we record some details of the derivation of the coefficients in 
the plateau model (\ref{eq:td1})--(\ref{eq:td3}).
We begin by noting that
$wu_w=\tsfrac12 f \eta\phi'(\eta)+\nu\phi_\nu=\tsfrac12 f(\eta\phi'+w\nu\psi)$,
$xu_x=\tsfrac12f\eta\phi'(\eta)$ and $u_{xx}=\tsfrac12fw^2\phi''(\eta)$.

Now we list the various inner products that are needed for the calculation:
\bean
\ave{u_f^2}&=&\tsfrac14\ave{\phi^2}\\
\ave{u_wu_f}&=&\frac{f}{4w}\ave{(\eta\phi'+w\nu\psi)\phi}
=-\frac{f}{8w}\ave{\phi^2}+\frac{f\nu}{4}\ave{\psi\phi}\\
\ave{u_\nu u_f}&=&\tsfrac14 fw\ave{\psi\phi}\\
\ave{\eta\phi' u_f}&=&\tsfrac12 \ave{\eta\phi'\phi}=-\tsfrac14\ave{\phi^2}\\
\ave{\phi'' u_f}&=&\tsfrac12\ave{\phi''\phi}=-\tsfrac12\ave{{\phi'}^2}\\
\ave{u(\alpha-u)(u-1)u_f}&=&\tsfrac14 f\left[-\alpha\ave{\phi^2}+
\tsfrac12(1+\alpha)f\ave{\phi^3}-\tsfrac14f^2\ave{\phi^4}\right]\\
\ave{u_w^2}&=&\frac{f^2}{4w^2}\left[\ave{\eta^2{\phi'}^2}
              +2w\nu\ave{\eta\phi'\psi}+w^2\nu^2\ave{\psi^2}\right]\\
\ave{u_\nu u_w}&=&\tsfrac14 f^2\ave{\eta\phi'\psi}+
                  \tsfrac14 f^2w\nu\ave{\psi^2}\\
\ave{\eta\phi' u_w}&=&\frac{f}{2w}\ave{\eta^2{\phi'}^2}+
\frac{f\nu}{2}\ave{\eta\phi'\psi}\\
\ave{\phi''u_w}&=&\frac{f}{2w}\ave{\phi''(\eta \phi'+w\nu\psi)}=
-\frac{f}{4w}\ave{{\phi'}^2} -\frac{fa}{4w}\pder{}{a}\ave{{\phi'}^2}\\
\ave{u(\alpha-u)(u-1)u_w}&=&
\frac{f^2}{4w}\left(a\pder{}{a}-1\right)
\left[-\tsfrac12\alpha \ave{\phi^2}+\tsfrac16(1+\alpha)f\ave{\phi^3}
-\tsfrac1{16}f^2\ave{\phi^4}\right]\\
\ave{u_\nu^2}&=&\tsfrac14 f^2w^2\ave{\psi^2}\\
\ave{\eta\phi' u_\nu}&=&\tsfrac12fw\ave{\eta\phi'\psi}\\
\ave{\phi''u_\nu}&=&-\tsfrac14 fw\pder{}{a}\ave{{\phi'}^2}\\
\ave{u(\alpha-u)(u-1)u_\nu}&=&\tsfrac14f^2w\pder{}{a}
\left[-\tsfrac12\alpha\ave{\phi^2}+\tsfrac16(1+\alpha)f\ave{\phi^3}
-\tsfrac1{16}f^2\ave{\phi^4}\right].
\eean

All of the requisite inner products can be found exactly:
\bean
\ave{\phi^2}&=&4(2a\coth2a -1)\\
\ave{\psi\phi}&=&\frac{1}{2}\pder{}{a}\ave{\phi^2}=
4\coth2a-8a\cosech^22a\\
\ave{{\phi'}^2}&=&\tsfrac83(1+3\cosech^22a-6a\cosech^22a \coth2a)\\
\ave{\phi^3}&=&4(4a\coth^22a+2a\cosech^22a-3\coth2a)\\
\ave{\phi^4}&=&\tsfrac83(-11+12a\coth^32a-15\cosech^22a+18a\coth2a\cosech^22a)\\
\ave{\eta\phi'\psi}&=&-\tsfrac83a\\
\ave{\psi^2}&=&\tsfrac83(1-3\cosech^22a+6a\cosech^22a\coth2a)\\
\ave{\eta^2{\phi'}^2}&=&
2\left(\tsfrac19\pi^2-\tsfrac{2}{3}+\tsfrac43a^2\right)-
\tsfrac43a(\pi^2+4a^2)\coth2a\cosech^22a+\tsfrac23(\pi^2+12a^2)\cosech^22a.
\eean

The various coefficients in  (\ref{eq:td1})--(\ref{eq:td3}) are then given by
\bean
&&\lambda_1=\tsfrac14\ave{\phi^2}\; ,\quad 
\mu_1=\lambda_2=-\tsfrac18\ave{\phi^2}+\tsfrac14 a\ave{\psi\phi}\; ,\quad
\beta_1=\lambda_3=\tsfrac14\ave{\psi\phi}\; ,\\
&&\lambda_4=-\tsfrac18\ave{\phi^2}\; ,\quad
\lambda_5=-\tsfrac14\ave{{\phi'}^2}\; ,\quad
\lambda_6=-\tsfrac14\alpha\ave{\phi^2}\; ,\quad
\lambda_7=\tsfrac18(1+\alpha)\ave{\phi^3}\; ,\quad
\lambda_8=-\tsfrac1{16}\ave{\phi^4}\; ,\\
&&\mu_2=\tsfrac14\left[\ave{\eta^2{\phi'}^2}
              +2a\ave{\eta\phi'\psi}+a^2\ave{\psi^2}\right]\; ,\quad
\beta_2=\mu_3=\tsfrac14\left[\ave{\eta{\phi'}\psi}+a\ave{\psi^2}\right]\; ,\\
&&\mu_4=\tsfrac14\ave{\eta^2{\phi'}^2}+\tsfrac14 a\ave{\eta\phi'\psi}\; ,\quad
\mu_5=-\tsfrac18\left(1+a\pder{}{a}\right)\ave{{\phi'}^2}\; ,\quad
\mu_6=\tsfrac18\alpha\left(1-a\pder{}{a}\right)\ave{\phi^2}\; ,\\
&&\mu_7=-\tsfrac{1}{24}(1+\alpha)\left(1-a\pder{}{a}\right)\ave{\phi^3}\; ,\quad
\mu_8=\tsfrac{1}{64}\left(1-a\pder{}{a}\right)\ave{\phi^4}\; ,\\
&&\beta_3=\tsfrac14\ave{\psi^2}\; ,\quad
\beta_4=\tsfrac14\ave{\eta\phi'\psi}\; ,\quad
\beta_5=-\tsfrac18\pder{}{a}\ave{{\phi'}^2}\; ,\\
&&\beta_6=-\tsfrac18\alpha\pder{}{a}\ave{\phi^2}\; ,\quad
\beta_7=\tsfrac{1}{24}(1+\alpha)\pder{}{a}\ave{\phi^3}\; ,\quad
\beta_8=-\tsfrac{1}{64}\pder{}{a}\ave{\phi^4}.
\eean


\medskip

{\underbar{\bf Acknowledgements }} G.A.G gratefully acknowledges
support by the Australian Research Council, DP0452147. 
\vskip 5pt
\vfill\eject




\clearpage

\begin{thebibliography}{99}

\bibitem{Epstein95}
I.~R.~Epstein, {The consequences of imperfect mixing in autocatalytic
chemical and biological systems}, {Nature} {374} (1995) 321--327.

\bibitem{Allen96}
M.~A.~Allen, J.~Brindley, J.~Merkin and M.~J.~Piling, {Autocatalysis
in a shear flow},
{Phys. Rev. E} {54} (1996)  2140--2142.

\bibitem{Abraham98}
E.~R.~Abraham, {The generation of plankton patchiness by turbulent
stirring}, {Nature} {391} (1998)  577--580.

\bibitem{Martin00}
A.~P.~Martin, {On filament width in oceanic plankton distributions},
{J.~Plank.~Res.} {22} (2000)  597--692.

\bibitem{McLeod02}
P.~McLeod, A.~P.~Martin and K.~J.~Richards, {Minimum length scale for
growth-limited oceanic plankton distributions}, {Ecological Modelling}
{158} (2002)  111--120.

\bibitem{HG04}
E.~Hern{\'a}ndez-Garcia and C.~Lop{\'e}z, {Sustained plankton
blooms under open chaotic flows}, {Ecological Complexity}
{1} (2004)  193.

\bibitem{Edouard96}
S.~Edouard, B.~Legras, F.~Lef{\`e}vre and R.~Eymard, {The effect of
small-scale inhomogeneities on ozone depletion in the Arctic},
{Nature} {384} (1996)  444--447.

\bibitem{Ronney94}
P.~D.~Ronney,
{Some open issues in premixed turbulent combustion}, in {\emph{Modeling
in Combustion Science}}, pp. 3--22, Eds. J.~Buckmaster and T.~Takeno
(Springer-Verlag Lecture Notes in Physics, 1994)

\bibitem{Neufeld01}
Z.~Neufeld, {Excitable media in a chaotic flow},
{Phys. Rev. Lett.} {87} (2001)  108301--108304.

\bibitem{Neufeld02}
Z.~Neufeld, P.~H.~Haynes and T.~T\'{e}l, {Chaotic mixing induced transitions in
reaction--diffusion systems}, {Chaos} {12} (2002)  426--438.

\bibitem{Neufeld02b}
Z.~Neufeld, C.~Lop{\'e}z, E.~Hern{\'a}ndez-Garcia and O.~Piro,
{Excitable media in open and closed chaotic flows},
{Phys. Rev. E} {66} (2002)  066208--066220.

\bibitem{HG03}
E.~Hern{\'a}ndez-Garcia, C.~Lop{\'e}z and Z.~Neufeld, {Filament
bifurcations in a one-dimensional model of reacting excitable fluid
flow}, {Physica A} {327} (2003)  59--64.

\bibitem{Kiss03b}
I.~Z.~Kiss, J.~H.~Merkin and Z.~Neufeld, {Combustion initiation and
extinction in a 2D chaotic flow}, {Physica D} {183}
(2003)  175--189.

\bibitem{Kiss03}
I.~Z.~Kiss, J.~H.~Merkin, S.~K.~Scott, P.~L.~Simon, S.Kalliadasis and
Z.~Neufeld, {The structure of flame filaments in chaotic flows},
{Physica D} {176} (2003)  67--81.

\bibitem{Menon05}
S.~Menon and G.~A.~Gottwald,
{Bifurcations in reaction--diffusion systems in chaotic flows},
{Phys. Rev. E} {\bf 71} (2005) 066201--066207.

\bibitem{Ranz79}
W.~E.~Ranz, 
{Applications of a stretched model to mixing, diffusion, and reaction
in laminar and turbulent flows},
AIChE J. \textbf{25} (1979) 41--47.

\bibitem{Gray94}
P.~Gray and S.~K.~Scott,
{\it Chemical Oscillations and Instabilities}, Clarendon Press, Oxford (1994).

\bibitem{r79}
W.~E.~Ranz,
{Applications of a stretch model to mixing, diffusion, and reaction
in laminar and turbulent flows},
{AIChE J.} {25} (1979)  41--47.

\bibitem{Boozer99}
X.~Z.~Tang and A.~H.~Boozer, {Design criteria of a chemical reactor
based on a chaotic flow}, {Chaos} {9} 183--194 (1999).

\bibitem{JLT03}
J.--L.~Thiffeault,
{Advection--diffusion in Lagrangian coordinates},
{Phys. Lett. A} {309} 415--422 (2003).

\bibitem{Cox04}
S.~M.~Cox, {Chaotic mixing of a competitive--consecutive reaction},
{Physica D} {199} (2004)  369--386.

\bibitem{Murray}
J.~D.~Murray, {\it Mathematical Biology}, 2nd edition, Springer
Verlag, New York (1993).

\bibitem{Keener} J.~Keener and J.~Sneyd, {\it Mathematical Physiology},
Springer Verlag, New York (1998).

\bibitem{Mikhailov} 
A.~S.~Mikhailov, {\it Foundations of synergetics I: Distributed active
systems}, 2nd edition, Springer Verlag, Berlin (1994).

\bibitem{Scott} A.~Scott, {\it Nonlinear Science -- Emergence and
Dynamics of Coherent Structures}, Oxford University Press, Oxford (1999).

\bibitem{GottwaldKramer04}
G.~A.~Gottwald and L.~Kramer, {On propagation failure in one- and
two-dimensional excitable media}, {Chaos} {14} (2004)  855--863.

\bibitem{Vikhansky04}
A.~Vikhansky,
{Quantification of reactive mixing in laminar microflows},
{Phys. Fluids} {16} (2004)  4738--4741.

\bibitem{CV06}
A.~Vikhansky and S.~M.~Cox,
{Reduced models of chemical reaction in chaotic flows},
{Phys. Fluids} {18} (2006) .

\end{thebibliography}
\end{document}